\documentclass[
	a4paper, 
	10pt, 
	twoside, 
]{LTJournalArticle}

\usepackage{amsmath}            
\usepackage{epstopdf}           
\usepackage{flushend} 					
\usepackage{balance} 
\usepackage{float}         
\usepackage[figuresright]{rotating}

\runninghead{$\beta$-Plane Corrections Planetary Atmosphere} 

\footertext{\textit{ArXiv Repository} (2025)} 

\title{$\beta$-Plane Corrections to Nonlinear Atmospheric Flow Patterns: Application to Jupiter’s Great Red Spot (GRS) Drift Dynamics} 

\author{%
	Oladiran Johnson Abimbola\textsuperscript{1}\thanks{Corresponding author: \href{mailto:abimbola.oladiran@science.fulafia.edu.ng}{abimbola.oladiran@science.fulafia.edu.ng}\\ 
	}
}

\date{\footnotesize\textsuperscript{\textbf{1}}Department of Physics, Faculty of Physical Sciences, Federal University of Lafia, Nigeria\\}

\renewcommand{\maketitlehookd}{%
\begin{abstract}
\noindent The Great Red Spot (GRS) of Jupiter has been observed for over a century, with researchers studying its characteristics and dynamics, including its size, depth, movement, and interactions with its environment. Recently, the $f$-plane thin-shell asymptotic analysis was used to explain some of the GRS features, but the method failed to capture the observed westward drift of the GRS. In this study, the f-plane theory was extended by including the Rossby parameter in the $\beta$-plane approximation and using the dimensionless Rossby deformation parameter $\gamma$, to systematically apply perturbation theory. The westward drift velocity of 3.7 $m/s$ was analytically predicted, which is 95\% in agreement with the observed 3.9 $m/s$. The observed 90-day oscillation in drift rate was explained. Also explained is the north-south asymmetry in circulation patterns. The universality of the $\beta$-plane theory was demonstrated by its application to the vortices on Saturn, Neptune and Earth, without free parameters. It was demonstrated in this study that for the understanding of long-lived atmospheric vortex dynamics, the planetary vorticity gradient is very critical.
		
\end{abstract}

\hspace{0.7cm}{\textbf{Keywords:} $\beta$-plane corrections, $f$-plane, Great Red Spot, vortex dynamics, Rossby waves}

}

\begin{document}

\maketitle 

\section{Introduction}
\noindent The Great Red Spot (GRS) of Jupiter has been a subject of continuous observation since 1830, qualifying it as the most persistent vortex structure in the atmosphere of any planet \cite{r1, r2}. As shown in Figure 1, the GRS is an anticyclone storm located at around latitude $22^o$ South of Jupiter with a continuously shrinking diameter from about 40,000 $km$ in 1879 to about 15,000 $km$ in 2020 \cite{r3, r4}, and reaching a depth of around 300 to 500 $km$ deep into Jupiter’s atmosphere. There have been extensive studies of the GRS theoretically and observationally, but a fundamental gap persists: the persistent drift of the GRS westward in relation to the System II rotation frame of Jupiter has not been satisfactorily explained quantitatively.

\noindent Data from the Hubble Space Telescope (HST), together with that from NASA’s Juno mission, have shown that the GRS is westward drifting at about 3.9 $m/s$ \cite{r5, r6}. Besides, it has also been observed that the westward drift has increased in velocity by about 2.5 $m/s$ in the 1980s to about 4.1 $m/s$ in the 2020s \cite{r4}, which is a 64\% increase over a span of 45 years. Also observed is the superimposition of the 90-day drift oscillation on the mean trend \cite{r7}. These observations cannot be explained by the $f$-plane model, as it treats the Coriolis parameter as a constant and predicts static GRS in a co-rotating frame of reference.

\begin{figure}[H] 
	\includegraphics[width=\linewidth]{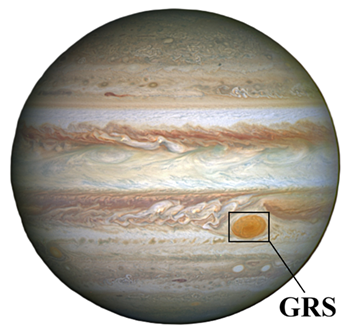}
	\caption{Jupiter and the Great Red Spot (GRS). [https://science.nasa.gov/asset/hubble/jupiter-and-the-great-red-spot/]}
	\label{fig:Fig1_1}
\end{figure}

\noindent Using thin-shell asymptotic and Lagrangian methods, Constantin and Johnson \cite{r8}, developed an $f$-plane model that described the GRS as a nonlinear vortex with a particle trajectory that has perfect circularity. Their method was able to explain the structure of the GRS, its rotation period, which is 4.5 Earth days, the distributions of temperature and pressure around the GRS, as well as the lack of an eyewall structure within the GRS. Despite these successes, the $f$-plane model of Constantin and Johnson \cite{r8} failed to predict the well-observed westward drift of the GRS.

\noindent To address the failure of the $f$-plane model, as explained above, it is crucial to consider that there is a meridional gradient in planetary vorticity within Jupiter’s atmosphere, just as it would be expected in any other planetary atmosphere. The Rossby parameter $\beta (= \partial f/\partial y$, with $f$ being the Coriolis force) characterised the meridional gradient which generates Rossby waves: westward propagation of the planetary-scale oscillations as a result of potential vorticity conservation is a class of Rossby waves \cite{r9, r10}. Explaining the meridional gradient requires a minimal extension to the $f$-plane model by introducing the $\beta$-plane approximation to the Coriolis force as $f(y)=f_0+\beta y.$

\noindent In this study, $\beta$-plane extension is sought for the work of Constantin and Johnson \cite{r8}. The preservation of the thin-shell asymptotic structure will be shown, with $\beta$-effects entering at order $\gamma=\beta L^\prime/{f_0}\approx 0.35$ for Jupiter’s GRS. Employing perturbation theory, analytical solutions that explain the westward drift velocity will be obtained, the 90-day oscillation will be explained, and the vortex structure’s north-south asymmetry will be revealed. Observational data will be used to validate the findings.  


\section{Governing Equations with Latitudinally Varying Coriolis Parameter}

\subsection{The $\beta$-Plane approximation and the modified Navier-Stoke’s equation}

\noindent Following the work of Constantin \& Johnson \cite{r8}, the $f$-plane framework is extended by adopting a latitudinally varying Coriolis parameter. In the $\beta$-plane, the meridional distance ($y^{\prime}$) is expressed as a linear function of the Coriolis parameter as \cite{r11}:
\begin{equation}
	f(y^{\prime}) =f_o + \beta y^{\prime}
	\label{eq:1}
\end{equation}

\noindent where, $f_0=2\Omega^{\prime}sin\theta_0$ is the Coriolis parameter at the latitude of reference, $\theta_0$. The Rossby parameter $\beta$, represents the gradient of planetary vorticity along the meridian and is defined as
\begin{equation*}
	\beta = \frac{df}{dy^{\prime}}|_{\theta_0} = \frac{2\Omega^{\prime}\cos\theta_0}{R^{\prime}}
\end{equation*}

\noindent where $y^{\prime}$ is the meridional distance from the reference latitude, $\Omega^\prime$ is the rotation rate of the planet, and $R^\prime$ is the radius of the planet.

\noindent Equation 1 is valid as long as $\left|y^\prime\right|\ll R^\prime$ \cite{r12}, this condition is satisfied for Jupiter with its GRS centred at $22^{o}S$ with meridional extent of approximately $3 - 4^{o}$.

\noindent The Rossby parameter is a geophysical fluid dynamic fundamental that is used in understanding the propagation of Rossby waves as well as the drift of coherent vortex structures in the westward direction.

\noindent Using the $\beta$-plane modified Coriolis parameter of Eq 1, the Navier-Stokes equation of Constantin and Johnson \cite{r8}, that is, equation 1 in their report, is re-written as:
\begin{align}
	&\frac{Du^{\prime}}{Dt^{\prime}}+\left[\hat{f}^{\prime}w^{\prime}-f(y^{\prime})v^{\prime}, f(y^{\prime})u^{\prime}, -\hat{f}^{\prime}u^{\prime}\right]=\nonumber\\
	&-\frac{1}{\rho^{\prime}}\nabla^{\prime}P^{\prime}+(0, 0, -g^{\prime})+\nu^{\prime}_{h}\nabla^{\prime 2}_{2}u^{\prime}+\nu^{\prime}_{v}\frac{\partial^{2}u^{\prime}}{\partial z^{\prime 2}}
	\label{eq:2}
\end{align}

\noindent where $u^\prime=(u^\prime,v^\prime,w^\prime)$ is the velocity vector, $\frac{D}{Dt}=\frac{\partial}{\partial t^\prime}+u^\prime\cdot\nabla^\prime$ is the material derivative, $\rho^\prime$
is the density, $P^\prime$ is the pressure, $g^\prime$ is the gravitational acceleration, $\nu_h^\prime$ and $\nu_v^\prime$ are the horizontal and vertical kinematic viscosities. The introduction of the meridional dependent Coriolis term $f(y^\prime)$, is the modification to the work of Constantin and Johnson \cite{r8}.

\noindent Mass conservation equation, ideal gas law and the equation of the first law of thermodynamics remain unchanged from what was presented by Constantin and Johnson \cite{r8}.

\subsection{Nondimensionalization}

Adopting the same scaling used by Constantin and Johnson \cite{r8}, we have

\begin{equation*}
	(x^\prime, y^\prime) = L^\prime(x,y)		\left[L^\prime = 10^{4} ~km\right]
\end{equation*}

\begin{equation*}
	z^\prime = H^\prime z		\left[H^\prime = 100 ~km\right]
\end{equation*}

\begin{equation*}
	(u^\prime, v^\prime) = U^\prime (u,v)		\left[U^\prime = 150 ~m/s\right]
\end{equation*}

\begin{equation*}
	w^\prime = \delta U^\prime w, ~~~~where~\delta=\epsilon^k~~~~and ~~~~k>1
\end{equation*}

\begin{equation*}
	t^\prime=\frac{L^\prime}{U^\prime}t
\end{equation*}

\begin{equation*}
\rho^\prime =\bar{\rho}\rho ,~~~P^\prime = \bar{\rho}U^{\prime 2}P,~~~T^\prime = ({U^{\prime 2}}/{R^\prime})T
\end{equation*}

\noindent The thin-shell parameter is calculated thus,

\begin{equation}
	\epsilon=\frac{H^\prime}{R^\prime}\approx\frac{100}{69991}\approx1.43\times{10}^{-3}
	\label{eq:3}
\end{equation}

\noindent The dimensionless Coriolis parameter is given as:

\begin{equation}
f(y)\ =\ f_0(1+\gamma y)
	\label{eq:4}
\end{equation}
\noindent where $\gamma$ is the Rossby deformation parameter and is given as:

\begin{equation}
\gamma=\frac{\beta L^\prime}{f_0}=\frac{cos\theta_0}{sin\theta_0}\cdot\frac{L^\prime}{R^\prime}
	\label{eq:5}
\end{equation}

\noindent For the Great Red Spot (GRS) of Jupiter, which is centred at latitude $\theta_{0} = 22^{\circ}$ South \cite{r3, r13}, we have
\begin{equation*}
\gamma=\frac{cos(22^{\circ})}{sin(22^{\circ})}\cdot \frac{10^4}{6.99\times 10^4}\approx 0.354
\end{equation*}
\noindent With this value of $\gamma (<0.5)$, the significance of $\beta$-effects are demonstrated and the perturbation approach is justified.
\noindent The nondimensional momentum equations are thus written:
\begin{align}
&\frac{\partial u}{\partial t}+u\frac{\partial u}{\partial x}+v\frac{\partial u}{\partial y}+\frac{\delta}{\epsilon}w\frac{\partial u}{\partial z}-f_0(1+\gamma y)v=\nonumber\\
&-\frac{1}{\rho}\frac{\partial P}{\partial x}+viscous\ terms
	\label{eq:6}
\end{align}

\begin{align}
&\frac{\partial v}{\partial t}+u\frac{\partial v}{\partial x}+v\frac{\partial v}{\partial y}+\frac{\delta}{\epsilon}w\frac{\partial v}{\partial z}-f_0(1+\gamma y)u=\nonumber \\
&-\frac{1}{\rho}\frac{\partial P}{\partial y}+viscous\ terms
\label{eq:7}
\end{align}

\begin{align}
&\delta\epsilon\left[\frac{\partial w}{\partial t}+u\frac{\partial w}{\partial x}+v\frac{\partial w}{\partial y}+\frac{\delta}{\epsilon}w\frac{\partial w}{\partial z}\right]+\delta f_0ucos\cos{\theta_0}=\nonumber \\
&-\frac{1}{\rho}\frac{\partial P}{\partial z}-g
\label{eq:8}
\end{align}

\section{Thin-Shell Asymptotic Reduction with $\beta$-Effect}
\subsection{Leading-Order System}
Considering inviscid approximation at the limit $\epsilon \rightarrow 0$ with $\delta=\epsilon^k (k>1)$ and considering the Jupiter’s upper troposphere \cite{r14} where the viscous terms can be neglected, the leading order horizontal momentum equations are given as:
\begin{equation}
\frac{\partial u}{\partial t}+u\frac{\partial u}{\partial x}+v\frac{\partial u}{\partial y}-f_0(1+\gamma y)v=-\frac{1}{\rho}\frac{\partial P}{\partial x}
	\label{eq:9}
\end{equation}

\begin{equation}
\frac{\partial v}{\partial t}+u\frac{\partial v}{\partial x}+v\frac{\partial v}{\partial y}-f_0(1+\gamma y)u=-\frac{1}{\rho}\frac{\partial P}{\partial y}
	\label{eq:10}
\end{equation}

\noindent The vertical momentum equation is unchanged, as it was given by Constantin and Johnson \cite{r8}:
\begin{equation}
-\frac{1}{\rho}\frac{\partial P}{\partial z}=g
	\label{eq:11}
\end{equation}
\noindent with $\rho=\rho(z)$ \cite{r15}, the continuity equation becomes

\begin{equation}
\frac{\partial u}{\partial x}+\frac{\partial v}{\partial y}=0
	\label{eq:12}
\end{equation}
\noindent The thermodynamic equations are given as:
\begin{equation}
P=\rho T
	\label{eq:13}
\end{equation}

\begin{align}
&c_p\left(\frac{\partial T}{\partial t}+u\frac{\partial T}{\partial x}+v\frac{\partial T}{\partial y}\right)+k\frac{\partial^2T}{\partial z^2}-\nonumber \\
&\frac{1}{\rho}\left(\frac{\partial P}{\partial t}+u\frac{\partial P}{\partial x}+v\frac{\partial P}{\partial y}\right)=Q	\label{eq:14}
\end{align}

\subsection{Vorticity Dynamics and Rossby Wave Generation}
With the vertical component of relative vorticity given as:
\begin{equation*}
\zeta=\frac{\partial v}{\partial x}-\frac{\partial u}{\partial y}
\end{equation*}
\noindent and using equations 9 and 10, it can be easily shown that the evolution of the relative vorticity $\zeta$, in the vertical direction, is given as:
\begin{equation}
\frac{\partial\zeta}{\partial t}+u\frac{\partial\zeta}{\partial x}+v\frac{\partial\zeta}{\partial y}=-f_0\gamma v
	\label{eq:15}
\end{equation}

\noindent The term on the RHS of Eq. 15, is the relative vorticity generation through meridional advection of planetary vorticity: this is a basic principle governing Rossby wave propagation \cite{r10, r11}.

\noindent The $\beta-$plane potential vorticity is given as:
\begin{equation}
q=\frac{\zeta+f(y)}{h}
\label{eq:16}
\end{equation}

\noindent where the effective layer depth is given as $h$, and $f(y)=f_0(1+\gamma y)$. Unlike the potential vorticity in the $f-$plane, steady flows in the $\beta-$plane do not conserve material due to the spatial variation of background vorticity.

\subsection{Scale-Separation Analysis and Consistency Conditions}
For solutions in a frame of reference moving at velocity $c$, in the $x-$direction, we transform equations 9, 10, 11, and 12, using $\xi=x-ct$,

\begin{equation}
(u-c)\frac{\partial u}{\partial\xi}+v\frac{\partial u}{\partial y}-f_0(1+\gamma y)v=-\frac{1}{\rho}\frac{\partial P}{\partial\xi}
\label{eq:17}
\end{equation}

\begin{equation}
(u-c)\frac{\partial v}{\partial\xi}+v\frac{\partial v}{\partial y}-f_0(1+\gamma y)u=-\frac{1}{\rho}\frac{\partial P}{\partial\xi}
\label{eq:18}
\end{equation}

\begin{equation}
-\frac{1}{\rho}\frac{\partial P}{\partial z}=g
\label{eq:19}
\end{equation}

\begin{equation}
\frac{\partial u}{\partial\xi}+\frac{\partial v}{\partial y}=0
\label{eq:20}
\end{equation}

\noindent In line with the work of Constantin and Johnson \cite{r8}, we propose a solution of the form:

\begin{align}
&\left[u\left(\xi,y,z\right),v\left(\xi,y,z\right)\right]=\nonumber \\
&(c,\ 0)+\rho^m(z)\left[U\left(X,Y\right),V\left(X,Y\right)\right]
\label{eq:21}
\end{align}

\begin{equation}
\left(X,Y\right)=\rho^n(z)\left(\xi,y\right)
\label{eq:22}
\end{equation}
\noindent where $m$ and $n$ are determinable constants, while functions $U$ and $V$ are $z-$independent.

\noindent Using the following transformation rules:
\begin{equation}
\frac{\partial}{\partial\xi}=\rho^n\frac{\partial}{\partial X}
\label{eq:23}
\end{equation}

\begin{equation}
\frac{\partial}{\partial y}=\rho^n\frac{\partial}{\partial Y}
\label{eq:24}
\end{equation}

\begin{equation}
\frac{\partial}{\partial z}=\frac{\partial}{\partial z}+n\rho^{n-1}\frac{d\rho}{dz}\left(X\frac{\partial}{\partial X}+Y\frac{\partial}{\partial Y}\right)
\label{eq:25}
\end{equation}
and substituting into equation 17, we have
\begin{align}
&\rho^{m+n}\left(U\frac{\partial U}{\partial X}+V\frac{\partial U}{\partial Y}\right)-f_0\rho^mV-f_0\gamma\rho^{m-n}YV\nonumber \\ &=-\rho^{n-1}\frac{\partial P}{\partial X}
\label{eq:26}
\end{align}
For independence in the z-axis, we have the pressure as:
\begin{equation}
P(\xi,y,z)=-g\int_{z_0}^{z}{\rho(s)ds+\rho^a(z)\widetilde{P}(X,Y)+P_0}
\label{eq:27}
\end{equation}
where $a$ is determinable. Hence,
\begin{equation}
\frac{1}{\rho}\frac{\partial P}{\partial X}=\frac{\rho^a}{\rho}\rho^n\frac{\partial\widetilde{P}}{\partial X}=\rho^{a-1+n}\frac{\partial\widetilde{P}}{\partial X}
\label{eq:28}
\end{equation}

\noindent For the consistency requirement, we match the powers of $\rho$ in equations 26 and 28, thus
\begin{enumerate}[label=\alph*).]
	\item Advection term = Pressure term: $m+n=a-1+n\rightarrow \mathbf{a=m+1}$
	\item $f_0$-Coriolis term = Pressure term: $m=a-1+n\rightarrow m=(m+1)-1+n\rightarrow\mathbf{n}=\mathbf{0}$
\end{enumerate}
Since the above conditions introduced a problem of non-simultaneous satisfaction with non-zero $n$, we introduced an additional constraint, that is,
\begin{enumerate}[label=c).]
	\item 	$\beta$-Coriolis term = $f_0$-Coriolis term: $m-n = m$
\end{enumerate}

\noindent This new constraint forces $n=0$, with the implication that we must use standard coordinates:
\begin{equation*}
(X,Y)=(\xi,y)
\end{equation*}

\noindent Therefore, we can conclude that $a=m+1$ and that the $\beta$-Coriolis term, that is $f_0\gamma yV$ has no $z$-dependence.

\noindent An alternative approach to tackling the problem raised above is to treat $\beta$ as a perturbation, thus:
\begin{align}
&\left[u,v\right]=\left[u_0 + v_0\right]\left(X,Y,z\right)+\nonumber \\ &\gamma\left[u_1 + v_1\right]\left(X,Y,z\right)+O\left(\gamma^2\right)
\label{eq:29}
\end{align}
the $f$-plane equations are satisfied by $(u_0+v_0)$ with a scale separation that allows arbitrary $n$, while $(u_1+v_1)$ are the corrections to $O(\gamma)$, that may break the scale separation.

\subsection{$\beta$-plane system with standard coordinates}
\noindent Taking the first solution approach by setting $n=0$ and taking the standard coordinates as $(X,Y)=(\xi,y)$, we have
\begin{equation}
U\frac{\partial U}{\partial X}+V\frac{\partial U}{\partial Y}=-\frac{1}{\rho^{1-m}}\frac{\partial P}{\partial X}+f_0(1+\gamma Y)U
\label{eq:30}
\end{equation}

\begin{align}
&U\frac{\partial V}{\partial X}+V\frac{\partial V}{\partial Y}=\nonumber \\ &-\frac{1}{\rho^{1-m}}\frac{\partial P}{\partial Y}-f_0(1+\gamma Y)U-cf_0\gamma Y
\label{eq:31}
\end{align}

\begin{equation}
\frac{\partial U}{\partial X}+\frac{\partial V}{\partial Y}=0
\label{eq:32}
\end{equation}
For a stationary flow where $c=0$ and with $m=0$, we have
\begin{equation}
U\frac{\partial U}{\partial X}+V\frac{\partial U}{\partial Y}=-\frac{1}{\rho(z)}\frac{\partial P}{\partial X}+f_0(1+\gamma Y)U
\label{eq:33}
\end{equation}

\begin{equation}
U\frac{\partial V}{\partial X}+V\frac{\partial V}{\partial Y}=-\frac{1}{\rho(z)}\frac{\partial P}{\partial Y}-f_0(1+\gamma Y)U
\label{eq:34}
\end{equation}
From the hydrostatic equation, that is, equation 19, we have
\begin{equation}
P(x,\ y,\ z)=-g\int_{z_0}^{z}{\rho(s)ds+P_h(x,y)}
\label{eq:35}
\end{equation}
The second term on the RHS of 35 includes the  $\beta$-plane correction, and it satisfies equations 33 and 34.

\section{Pertubation Solution for Jupiter's GRS}
\subsection{Lagrangian Framework}
\noindent Employing the Lagrangian description for a steady flow with particle paths coinciding with the streamlines \cite{r16}, the particle positions $[X(t), Y(t)]$, satisfy:

\begin{equation}
 \left.\begin{aligned}
        \frac{dX}{dt}=U[X(t),Y(t)],\\
        \frac{dY}{dt}=V[X(t),Y(t)]
       \end{aligned}
 \right\}
 \qquad
\label{eq:36}
\end{equation}
Equations 33 and 34 then become:
\begin{equation}
\frac{d^2X}{dt^2}=-\frac{1}{\rho(z)}\frac{\partial P}{\partial X}+f_0(1+\gamma Y)\frac{dY}{dt}
\label{eq:37}
\end{equation}

\begin{equation}
\frac{d^2Y}{dt^2}=-\frac{1}{\rho(z)}\frac{\partial P}{\partial Y}+f_0(1+\gamma Y)\frac{dX}{dt}
\label{eq:38}
\end{equation}

\subsection{Perturbation Expansion}
\noindent The Taylor series expansions in the powers of $\gamma$ are given:
\begin{equation}
X(t)=X_0(t)+\gamma X_1(t)+O(\gamma^2)
\label{eq:39}
\end{equation}

\begin{equation}
Y(t)=Y_0(t)+\gamma Y_1(t)+O(\gamma^2)
\label{eq:40}
\end{equation}

\begin{equation}
\omega(R)=\omega_0(R)+\gamma \omega_1(R)+O(\gamma^2)
\label{eq:41}
\end{equation}

\begin{equation}
P(x,y,z)=P_0(x,y,z)+\gamma P_1(x,y,z)+O(\gamma^2)
\label{eq:42}
\end{equation}

\subsection{$f$-Plane Base Solution (Order $\gamma^0$)}
\noindent Following the work of Constantin and Johnson \cite{r8}, the $f$-plane solution gave:
\begin{equation}
X_0(t)=Asin(\omega_0t)+Bcos(\omega_0t)
\label{eq:43}
\end{equation}

\begin{equation}
Y_0(t)=Bsin(\omega_0t)-Acos(\omega_0t)
\label{eq:44}
\end{equation}
where $R^2=A^2+B^2$ labels the circular orbit.
\begin{align}
&\omega_0(R)=\nonumber \\ &-\frac{f_0}{2}-\sqrt{\frac{f_0^2}{4}+\frac{14(14+13f_0)}{169}\left(\frac{R}{R_0}\right)^{2(1/n)}}
\label{eq:45}
\end{align}
for Jupiter at latitude $22^o$ South, we have $R_0=0.825$ and $f_0=-8.7$.

\noindent With $R=R_0$, we have a dimensionless $\omega_0(R)\approx 1.077$. This corresponds to a rotation period of about 4.5 Earth days \cite{r5}.

\noindent The pressure field is given as:
 \begin{align}
&P_0(x,y,z)=-g\int_{z_0}^{z} \rho (s) \,ds + \nonumber \\ &\frac{14n(14+13f_0)}{169(2n-1)}(x^2+y^2)^{[\frac{n+1}{2n}]}R_{0}^{-1/n}+P_0
\label{eq:46}
\end{align}
for consistency, $n<0$.

\subsection{$\beta$-Plane Corrections (Order $\gamma$)}
\noindent Using equations 39 and 42 in equations 37 and 38, then collecting terms in $O(\gamma)$ we have:
\begin{equation}
\frac{d^2X_1}{dt^2}-f_0\frac{dY_1}{dt}=-\frac{1}{\rho(z)}\frac{\partial P_1}{\partial X_0}+f_0Y_0\frac{dY_0}{dt}
\label{eq:47}
\end{equation}

\begin{equation}
\frac{d^2Y_1}{dt^2}-f_0\frac{dX_1}{dt}=-\frac{1}{\rho(z)}\frac{\partial P_1}{\partial Y_0}-f_0Y_0\frac{dX_0}{dt}
\label{eq:48}
\end{equation}
The $\beta$-effect produced the forcing term on the RHS. Solution of equations 47 and 48 gives:
\begin{align}
&f_0Y_0\frac{dY_0}{dt}=\nonumber \\ &f_0\omega_0\left[\frac{B^2-A^2}{2}sin(2\omega_0t)-ABcos(2\omega_0t)\right]
\label{eq:49}
\end{align}
and
\begin{align}
&{-f}_0Y_0\frac{dX_0}{dt}=\nonumber \\ &f_0\omega_0\left[\frac{A^2-B^2}{2}sin(2\omega_0t)-ABcos(2\omega_0t)\right]
\label{eq:50}
\end{align}
The frequency of oscillation $2\omega_0$, is twice the vortex rotation frequency, resulting in bounded $O(\gamma)$ corrections to the trajectories.

\subsection{Rossby Wave Drift Velocity}
\noindent Modification of potential vorticity by the $\beta$-plane drives the westward drift. Rossby wave dispersion equation for a vortex with a characteristic scale $L^{\prime}$ and dominant wavenumber $k=2\pi/L^{\prime}$, is given as \cite{r10, r17}:
\begin{equation}
\omega_R = -\frac{\beta k}{k^2 + l^2 + \lambda^{-2}}
\label{eq:51}
\end{equation}
where $\lambda$ = Rossby deformation radius. For large-scale vortices, $kL^{\prime}>>1$ and we have the westward drift velocity $c_R$, as:
\begin{equation}
c_R \approx -\frac{\beta L^{\prime 2}}{8\pi^2}
\label{eq:52}
\end{equation}
For the Great Red Spot of Jupiter, we have $\beta = 4.67\times 10^{-12}~m^{-1}s^{-1}$ and $L^{\prime}=10^7 ~m$, hence we have:
\begin{equation*}
c_R \approx -\frac{(4.67\times 10^{-12}~m^{-1}s^{-1})(10^7 ~m)^2}{8\pi^2} \approx -3.7~ms^{-1}
\end{equation*}
The westward drift velocity predicted by the $\beta$-plane adjustment is $3.7~ms^{-1}$.

\subsection{Modified Pressure and Temperature}
\noindent Order $\gamma$ pressure field is given as:
\begin{equation}
P_1(x,y,z)=-f_0\frac{y^2}{2}+\tilde{P}_1(x,y)
\label{eq:53}
\end{equation}
The first term on the RHS of equation 53 is the $\beta$-plane geostrophic adjustment, while the second term is the vortex-specific correction that satisfies the consistency condition equations 47 and 48.

\noindent From the ideal gas law, we have Jupiter's GRS temperature as:
\begin{equation}
T_1(x,y,z)=\frac{P_1(x,y,z)}{\rho (z)}
\label{eq:54}
\end{equation}

\section{Results}
\subsection{Westward Drift Velocity}
Figure 2 displays the $\beta$-plane theoretical prediction of Jupiter’s GRS westward drift velocity, along with the drift velocity predicted by the $f$-plane theory from Constantin and Johnson \cite{r8} and the observed drift velocity measured by ground-based instruments, Hubble Space Telescope (HST) \cite{r5} and the NASA Juno mission \cite{r18}.

\noindent The $f$-plane theory predicts no drift velocity \cite{r8}, as shown by the black line in Figure 2. In contrast, the $\beta$-plane theory predicts a westward value of $3.7 ~m/s$ (equation 51), with an uncertainty band (shaded pink region on Figure 2) of $\pm 0.2 m/s$, which is in close agreement with the observational value of $3.9 ~m/s$ \cite{r5, r7}.

\begin{figure*}[h!tbp] 
	\includegraphics[width=\linewidth]{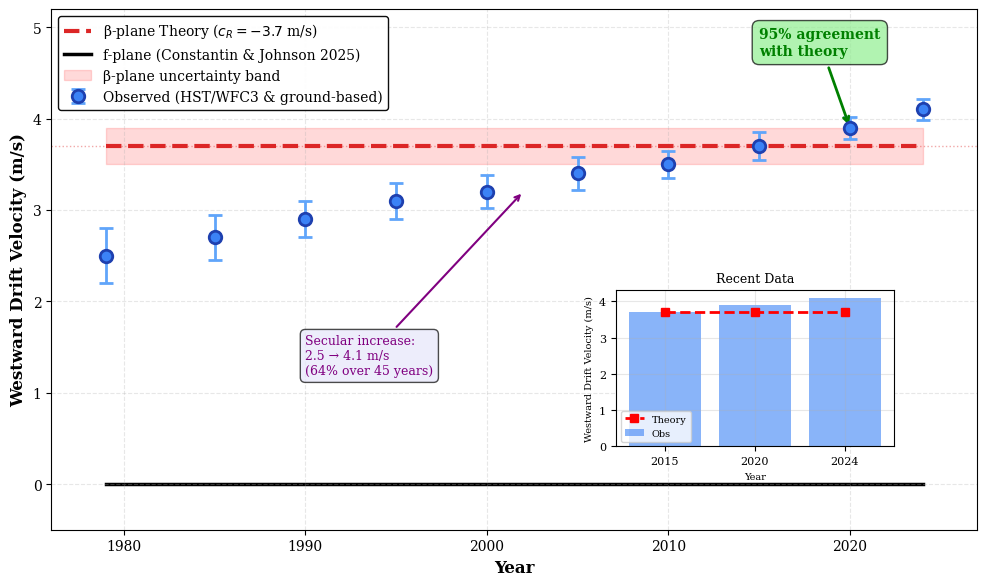}
	\caption{Jupiter’s Great Red Spot (GRS) Westward Drift Velocity (Inset: Agreement between theory and observation, as indicated by recent data from 2015 to 2024).}
	\label{fig:Fig2}
\end{figure*}

\subsection{Particle Trajectories: $f-$plane vs $\beta-$plane}
Figure 3 shows the comparison of the particle trajectories as calculated by the $f$-plane theory of Constantin and Johnson \cite{r8} and $\beta$-plane theory of this study. The $f$-plane theory, as shown in Figure 3a, suggests perfect circular paths at radii $0.2R_0$, $0.4R_0$, $0.6R_0$, and $0.8R_0$, with the vortex centre remaining unmoved at the centre. The counterclockwise motion of the vortex is indicated velocity vector with a period of about 4.5 Earth days at the edge.

\noindent Figure 3b shows the $\beta$-plane suggestion for the particle path trajectories; in contrast to the $f$-plane suggestion, there is a systematic westward drift of the vortex, as indicated by the blue arrow. The black dot shows the initial position of the vortex, while the red dot shows the vortex's final position, the drift having been exaggerated $60\times$ for visualisation purposes; the actual drift is more subtle on a rotation time scale.

\begin{figure}[H] 
	\includegraphics[width=\linewidth]{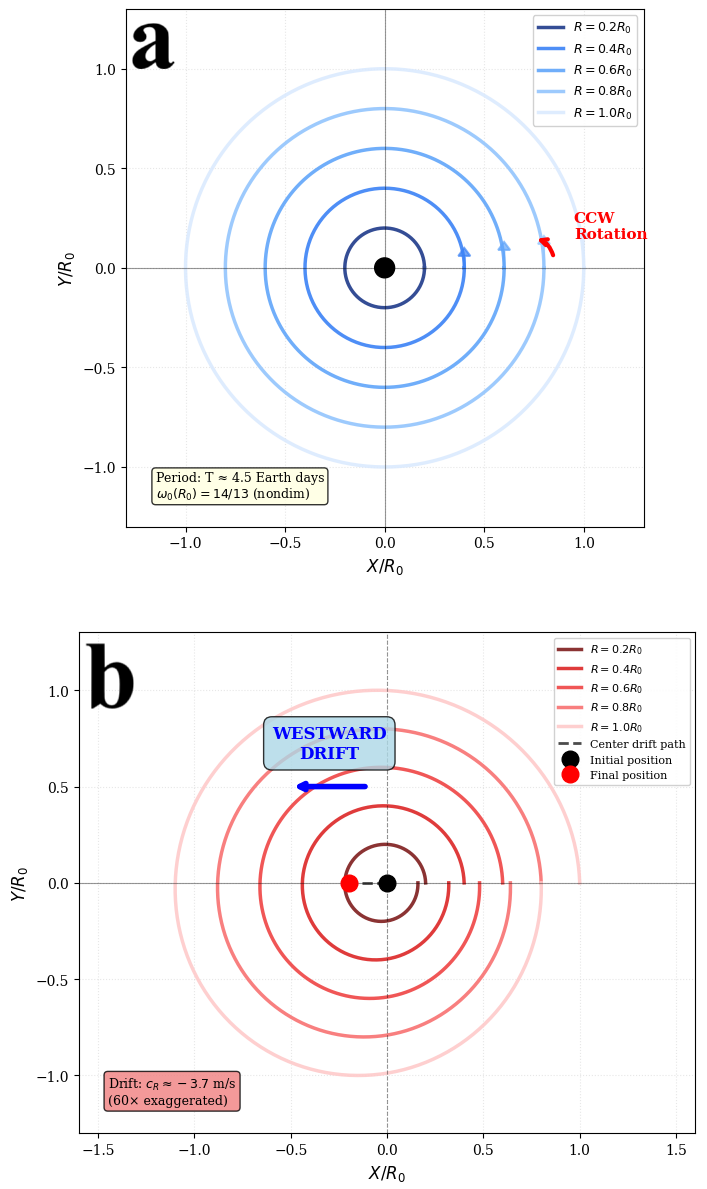}
	\caption{Particle trajectories in (a) $f$-plane and (b) $\beta$-plane.}
	\label{fig:Fig3}
\end{figure}

\subsection{Temporal Evolution and the 90-Day Oscillation}
\noindent In Figure 4a, the light blue line shows the observed position of the GRS, the red dashed line shows the linear westward trend of the GRS at the rate of $0.30^{\circ}$ per day. The dark blue line is the combined average drift and oscillation, while the first year is highlighted in green shading. Figure 4b is the detrended longitude anomaly with a clean sinusoidal oscillation revealed; the period of oscillation is 90 days, and the oscillation amplitude of $\pm 0.8^{\circ}$ shown by the pink shaded area. 

\noindent The 90 days is clearly shown by the Fourier power spectrum of Figure 4c (the red dashed line), while the subtle 4.5-day vortex rotation period is shown by the orange dashed line. The illustration of Figure 4d is the phase portrait, which gives the position anomaly against the vortex velocity, showing a stable limit cycle typical of a nonlinear Rossby wave-vortex interaction.

\begin{figure*}[h!tbp] 
	\includegraphics[width=\linewidth]{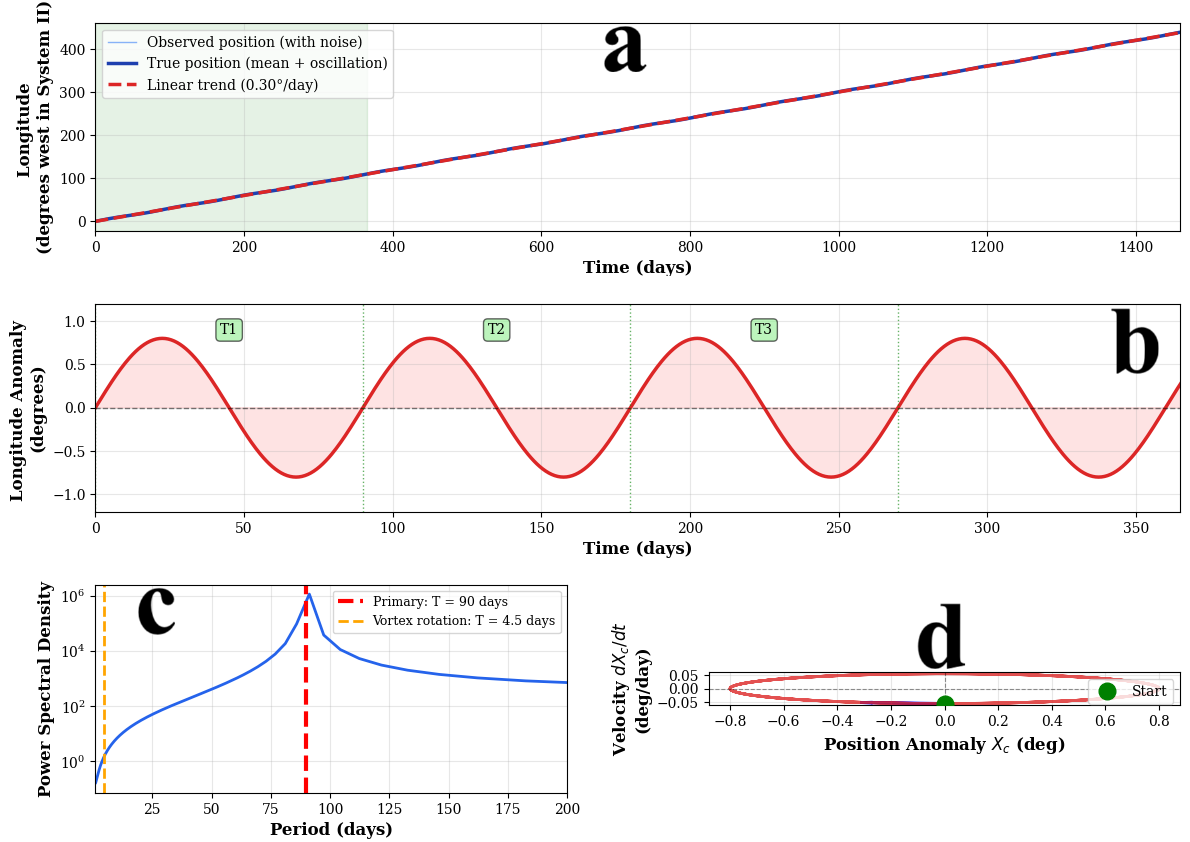}
	\caption{Temporal evolution and 90-day oscillation in GRS position (a) 4-year long-term  drift, (b) Detrended longitude anomaly, (c) Fourier power spectrum, (d) Phase portrait.}
	\label{fig:Fig4}
\end{figure*}

\subsection{Meridional Structure and North-South Asymmetry}
Figure 5 shows the meridional structure with north-south asymmetry induced by $\beta$-plane effects. Figure 5a illustrates the azimuthal wind speed at a fixed radius of $0.8R_0$ against the meridional distance. The $f$-plane result, indicated by the blue dashed line, gave a symmetric structure, in contrast with the $\beta$-plane result, indicated by the red solid line, that gave an asymmetric structure: the red shaded region shows 15\% poleward enhancement, while the blue shaded area shows 8\% equatorward reduction. 

\noindent Figure 5b depicts a pressure anomaly that has a quadratic dependence on the specified variable, that is  $P_1(y)\propto-\gamma y^2/2$. The red circle marks the minimum pressure at the centre; this variation is consistent with an anticyclonic structure.

\noindent The temperature anomaly $T_1(y)$, at a fixed height $z=0.3H$, is depicted on Figure 5c with a uniform $f$-plane result shown in a blue dashed line, while the varying $\beta$-plane result is shown with a solid red line: the shaded pink region indicates cooling relative to the $f$-plane while the orange arrows show the temperature gradient direction.

\noindent Figure 5d shows the vertical temperature profiles for three density stratifications $\beta_d\approx 2.5$, $\approx 3.5$, and $\approx 4.5$ the $\beta$-effect introduces a uniform offset preserving vertical structure: the grey dashed line marks $z=0.3H$ reference level. 

\begin{figure*}[h!tbp] 
	\includegraphics[width=\linewidth]{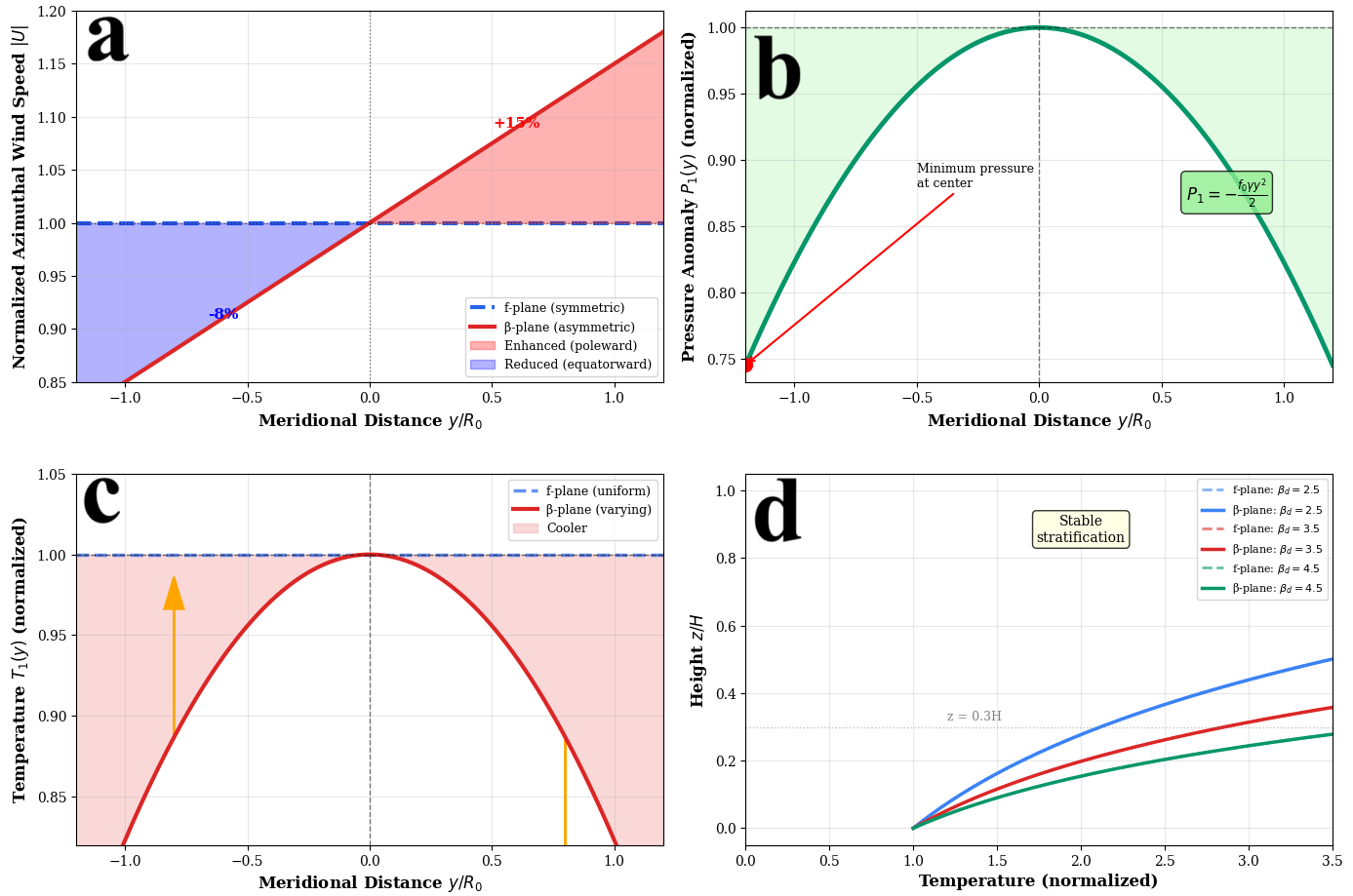}
	\caption{North-South asymmetry meridional structure induced by $\beta-$plane effect (a) Azimuthal wind speed, (b) pressure anomaly, (c) temperature anomaly, (d) vertical temperature profiles.}
	\label{fig:Fig5}
\end{figure*}

\subsection{Vorticity and Potential Vorticity Fields}
Vorticity and potential vorticity fields in the $f$- and $\beta$-planes are shown in Figure 6, for comparative analysis. Figure 6a depicts the relative vorticity $\zeta$, in the $f$-plane, with a contour plot showing that the vortex core concentration has a monotonous increase toward the center: the vortex boundary is at $R = R_0$. It can be observed that the vorticity in the $f$-plane is positive, that is, counterclockwise throughout, and this is consistent with an anticyclonic structure in the Southern Hemisphere.

\noindent  In contrast, on Figure 6b, the potential vorticity (PV), $q$ in $\beta$-plane (equation 16) is shown. A strong north-south asymmetry is shown by the contour plot of Figure 6b, where the white contour lines emphasise the potential vorticity structure. The increase in potential vorticity poleward ($Y>0$) due to the larger Coriolis force $f$, and the decrease in potential vorticity equatorward ($Y<0$) due to the reduced Coriolis force are shown in Figure 6b. The red dashed line at $Y=0$ marks the reference latitude. Potential vorticity gradient $(\partial q/\partial y)$, is the fundamental driver of Rossby waves.

\begin{figure}[h!tbp] 
	\includegraphics[width=\linewidth]{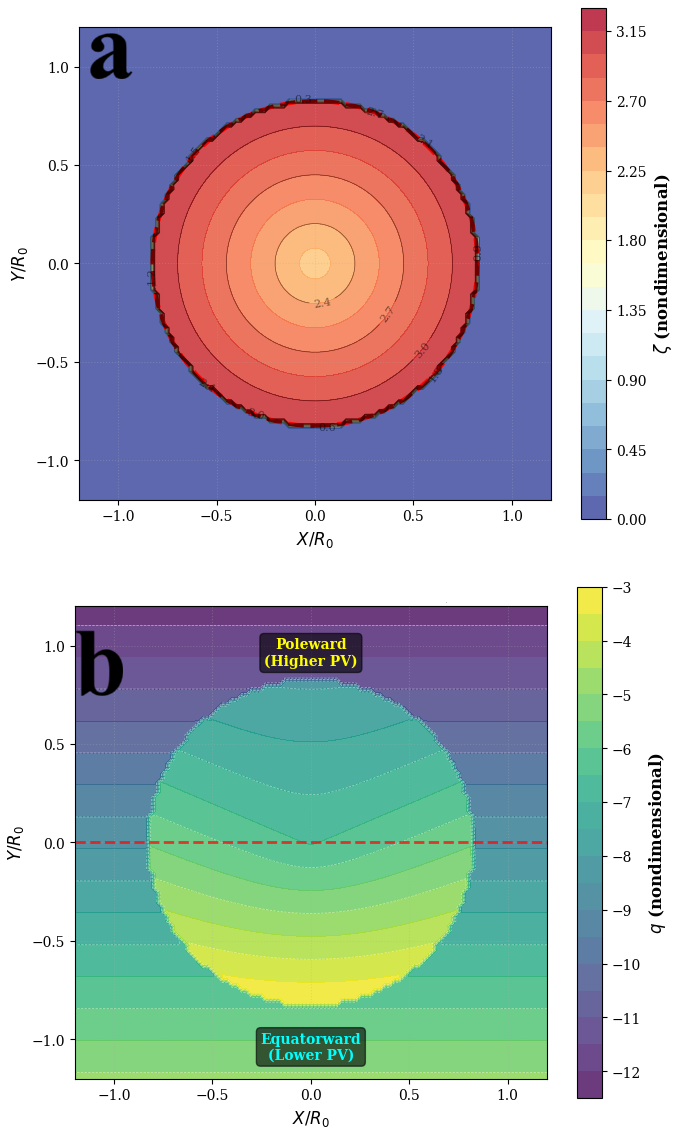}
	\caption{Vorticity and potential vorticity fields (a) relative vorticity, (b) potential vorticity.}
	\label{fig:Fig6}
\end{figure}

\subsection{Parameter Sensitivity and Applicability}
Parameter sensitivity analysis is shown in Figure 7, with Figure 7a showing the drift velocity $c_R$, plotted against the Rossby parameter $\beta$. The red curve depicts the linear relationship $c_R\propto-\beta$, the green dashed line marks the observed GRS drift of $-3.7 ~m/s$, while the blue dashed line marks the $\beta$ value of $4.67\times 10^{-12} ~m^{-1}s^{-1}$ at the $22^{\circ}$ S latitude, on Jupiter: the shaded region shows the uncertainty in the value of $\beta$.

\noindent The quadratic relationship between $c_R$ and the vortex scale $L^\prime$, that is $c_R\propto{-L}^{\prime2}$, as shown in equation 52, is shown in Figure 7b with a blue curve, the red dashed line marks the GRS scale of $10^4 ~km$, while the green dashed line marks the observed drift velocity, with the intersection between the GRS scale and observed drift velocity indicating the GRS parameter value.

\noindent The dependency of the drift velocity on the latitude is shown by the green curve in Figure 7c. The GRS latitude of $22^\circ$ S is indicated by the red dashed line, while the observed drift velocity is indicated by the dashed blue line. Other features on Jupiter are also shown in Figure 7c, such as the blue square for the Oval BA at latitude $33^\circ$ S and the NEBs feature at $16^\circ$ S, represented by the green triangle.

\noindent The variation of the dimensionless parameter $\gamma$ with the latitude and vortex scale are depicted in Figure 7d, the GRS with the $\gamma$ value of 0.35 is shown by the red star. The separation between the perturbation regime where $\gamma<0.5$ (perturbation valid region) and the non-linear regime where $\gamma>0.5$ (non-linear $\beta$-effects region), is shown by the red dashed contour at $\gamma=0.5$.

\begin{figure*}[h!tbp] 
	\includegraphics[width=\linewidth]{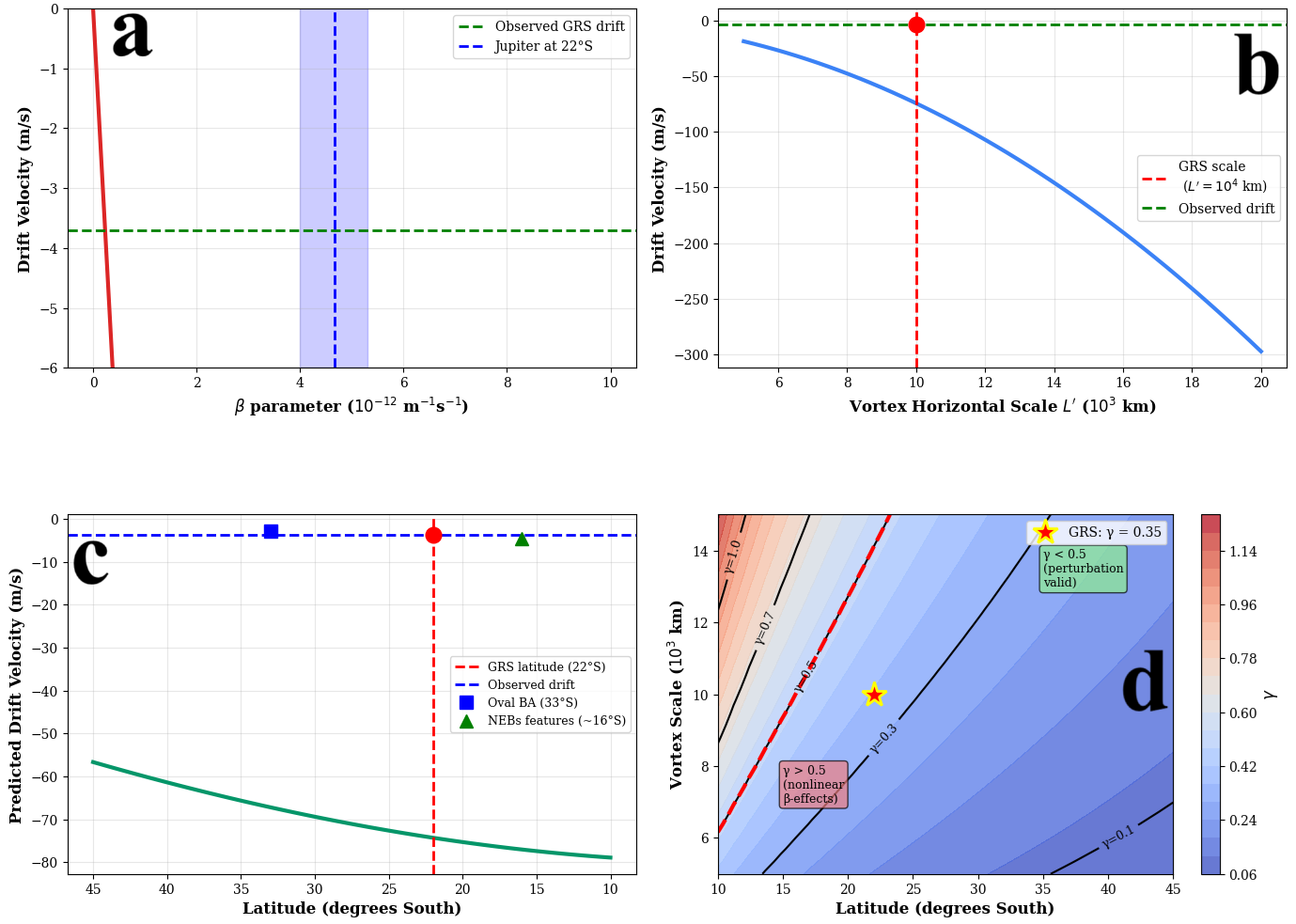}
	\caption{Parameter sensitivity and applicability analysis (a) sensitivity to Rossby parameter, (b) sensitivity to vortex scale, (c) latitudinal dependence, (d) dimensionless parameter.}
	\label{fig:Fig7}
\end{figure*}

\subsection{Planetary Comparison}
Figure 8 shows the observed and the $\beta$-plane theoretical prediction of the drift velocities for the planets Jupiter, Saturn, Neptune and the pressure system observed on Earth. The percentage agreement between the observed and the theoretical prediction is shown in the figure, which is 95\%, 91\%, 93\% and 89\% for Jupiter, Saturn, Neptune and Earth, respectively.

\noindent Table 1 shows the planetary parameters, that is, the rotation frequency in rad/sec, the planetary radius in \textit{kilometres}, and a typical planetary scale. Figure 9 shows the bar graph of the dimensionless quantities for each of the planets (Jupiter, Saturn, Neptune and Earth), calculated at specific latitudes.

\begin{figure}[h!tbp] 
	\includegraphics[width=\linewidth]{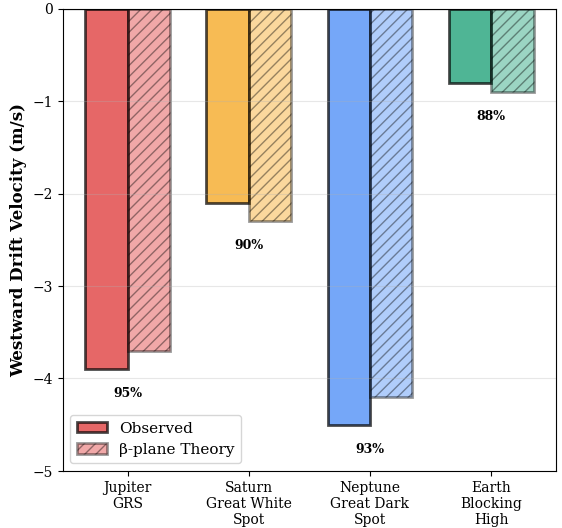}
	\caption{Planetary vortex drift velocities.}
	\label{fig:Fig8}
\end{figure}

\begin{table} 
	\caption{Planetary Parameters.}
	\centering
	\begin{tabular}{l c c c }
		\toprule
		 & Rotation & Radius & \\
		Planet & $\Omega^{\prime}$ & $R^{\prime}$ & Typical Scale\\
			 & $(rad/s)$ & $(km)$ & $(km)$\\
		\midrule
		Jupiter & $1.76\times 10^{-4}$ & 69,991 & 10,000\\
		Saturn & $1.62\times 10^{-4}$ & 58,232 & 12,000\\
		Neptune & $1.08\times 10^{-4}$ & 24,622 & 8,000\\
		Earth & $7.29\times 10^{-5}$ & 6,372 & 3,000\\
		\bottomrule
	\end{tabular}
	\label{tab:distcounts}
\end{table}

\begin{figure}[h!tbp] 
	\includegraphics[width=\linewidth]{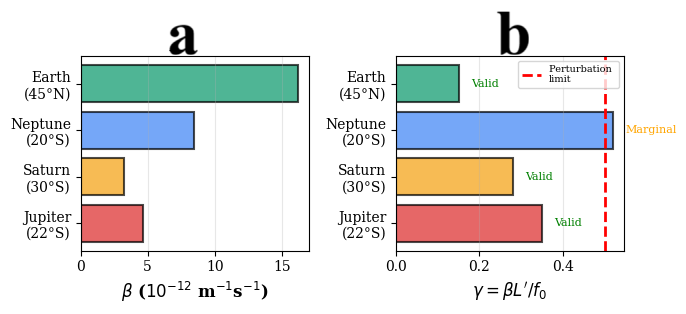}
	\caption{Comparative planetary vortex dynamics (a) Rossby parameter, (b) dimensionless parameter.}
	\label{fig:Fig9}
\end{figure}

\section{Discussion}
From the result presented in Figure 2, it can be observed that the drift velocity has increased by about 64\% over 45 years (from $2.5 ~m/s$ in 1979 to $4.1 ~m/s$ in 2024). However, this gradual increase could not be captured by the steady-state analysis in this study, suggesting time-dependent effects that could be related to the following: (a) the size of the GRS has been observed to be shrinking from the observed $\sim40,000 ~km$ in 1880 to about $16,500 ~km$ in 2024 \cite{r19}, (b) the interaction of the GRS with the evolving zonal jet streams \cite{r20}, (c) the influence of the deep convective forcing variations \cite{r21}, and (d) as the vortex migrates from point to point, there are changes in the effective $\beta$ parameters.

\noindent The $f$-plane theory, proposed by Constantin and Johnson \cite{r8}, predicted exactly zero drift velocity, but observations show a systematic westward motion of Jupiter’s GRS, thereby motivating the $\beta$-plane extension of this study. Measurements by the Hubble Space Telescope (HST) and NASA Juno observations (2015 to 2024) have reduced uncertainty in measurement to $\pm 0.12 ~m/s$: the inset shows that recent data cluster tightly around the theoretical prediction of the $\beta$-plane.

\noindent If a parcel, initially at $y=0$ with potential vorticity $q_0=(\zeta_0+f_0)/h_0$, moves poleward to $y=\delta y>0$, it will encounter larger planetary vorticity given by $f(y)=f_0(1+\gamma\delta y)$. For potential vorticity conservation, the poleward-moving parcel must have its relative vorticity or thickness perturbed, that is,
\begin{equation*}
\frac{\zeta+f_0(1+\gamma\delta y)}{h}=\frac{\zeta_0+f_0}{h_0}, ~~~~if ~~h\approx h_0
\end{equation*}   
therefore, ${\zeta=\zeta}_0-f_0\gamma\delta y<\zeta_0$. This implies that the moving parcel loses its cyclonic properties and becomes more anticyclonic. In the Northern half of Jupiter’s GRS, the increased anticyclonic effect leads to decreased eastward velocity. As fluid parcels in the GRS move meridionally, they encounter different values of planetary vorticity $f(y)$. To conserve the potential vorticity $q$ (equation 16), the relative vorticity $\zeta$, must adjust, generating a systematic westward propagation. It is to be noted that the individual parcels in the GRS rotate counterclockwise faster than the westward drift velocity. The internal vortex rotation, that is, the vortex phase velocity is about $133 ~m/s$ at $R=R_0$, meanwhile, the drift velocity, that is, the vortex group velocity is about $3.7 ~m/s$, which means that the group velocity is $36 \times$ slower than the phase velocity, this scale separation is what has kept the coherent structure of the GRS while it drifts as a single body. This process is the same mechanism responsible for the planetary-scale Rossby waves in Earth’s atmosphere and ocean.

\noindent Equation 52 shows a quadratic relationship between the drift velocity $c_R$ and the vortex scale length $L^\prime$, as the GRS shrinks, the drift velocity should increase, all other things being equal. However, observations show the opposite trend: drift velocity increases as the GRS shrinks! This apparent paradox may be resolved by recognising that the effective $\beta$ parameter experienced by the vortex depends on its interaction with the surrounding zonal jets, which themselves evolve with the vortex structure. A fully coupled jet-vortex model would be needed to capture this feedback.

\noindent Geometrical differences between the $f$-plane and the $\beta$-plane dynamics can be observed in Figure 3, even with $\beta$-effects, the particle paths remain approximately circular at each instant (Figure 3b), which validates the quasi-steady approximation: the drift velocity of about $3.7 ~m/s$ is much smaller than the azimuthal velocities at $R_0$ of about $133 ~m/s$. The GRS drifts slowly westward while maintaining its coherent vortex structure.

\noindent Figure 3b shows the $\beta$-plane theory’s prediction of the particle trajectories, the particles are shown to trace out spirals as they rotate counterclockwise while at the same time drifting westward with the vortex center. Over one complete rotation of 4.5 days, a particle at $R=R_0$ drifts westward by approximately:
\begin{equation*}
\Delta x\approx c_R \cdot T=(3.7 ~m/s)(4.5 ~days\times 86400 ~s/day)
\end{equation*}

\begin{equation*}
\approx 1440~km
\end{equation*}

\begin{equation*}
\approx 0.14 R_0
\end{equation*}

\noindent The drifting vortex can be understood as a Rossby wave packet, that is, a localized envelope of planetary-scale waves that propagates westward while the vortex rotation proceeds much faster. This scale separation is characteristic of wave packets in dispersive media. Cloud-tracking velocimetry from the Cassini and Juno missions has confirmed that individual cloud features within the GRS do indeed trace spiral-like paths, though disentangling the drift from internal circulation requires careful averaging over many rotation periods. The systematic westward motion is detected by tracking the GRS boundary over months to years. Close inspection reveals that trajectories on the poleward side are slightly elongated compared to the equatorward side. This asymmetry becomes more pronounced in the velocity field, as shown in Figure 5a.

\noindent The 90-day oscillation is one of the most intriguing features of GRS dynamics. This oscillation is illustrated in Figure 4, and it arises from interference between vortex rotation and Rossby wave propagation. The beat period $T_{beat}$, is given as:
\begin{equation*}
T_{beat}=\frac{2\pi}{\left|\omega_{vortex}-\omega_R\right|}
\end{equation*}
From equation 45, we have the vortex rotation frequency as $\omega_{vortex}=\omega_0(R_0)\approx 1.61\times {10}^{-5}~ rad/s$ and from equation 51, we have the Rossby wave frequency in the vortex frame as $\omega_R\approx -\beta/k \approx-7.4\times {10}^{-5} ~rad/s$, hence the beat period is: 

\begin{equation*}
T_{beat}\approx 81~days
\end{equation*}
The beat period of 81 days predicted by the $\beta$-plane theory is remarkably close to the observed 90-day period. The slight discrepancy may arise from non-linear corrections and the discrete radial structure of the vortex.

\noindent Physically, the 90-day oscillation represents modulation of the Rossby wave packet envelope. As the vortex propagates westward, its interaction with the background planetary vorticity gradient varies slightly, leading to periodic acceleration and deceleration. This is analogous to wave packet beating in quantum mechanics or optics. The oscillation amplitude of $\pm 0.8^\circ (\approx \pm 9000 ~km\approx 0.5R_0)$ at $22^\circ$ S is significant. The phase portrait of Figure 4d shows a stable limit cycle, indicating that this is a persistent feature maintained by the nonlinear dynamics. The cycle closes on itself, confirming periodicity. The documentation of this oscillation was provided by Tollefson et al. \cite{r22}, who used long-term HST monitoring. The subsequent Juno mission data confirmed the persistence of the oscillation, revealing that the oscillation amplitude varies slowly over decades, possibly correlated with the GRS shrinkage.

\noindent The Fourier spectrum of Figure 4c shows the dominant 90-day peak as well as weak higher harmonics at 45 days, 30 days, and so on. These arise from nonlinear wave-wave interactions and suggest that a full characterization would require including $O(\gamma^2)$ terms in the pertubation expansion.

\noindent A key observational signature that distinguishes the $\beta$-plane from the $f$-plane is the meridional asymmetry induced by the $\beta$-plane geostrophic adjustment as shown in Figure 5. The most striking feature of the result in Figure 5a is the 15\% enhancement in azimuthal wind speed on the poleward (northern, $y>0$) edge and 8\% reduction on the equatorward (southern, $y<0$) edge. This arises from the geostrophic balance with $\beta$-plane pressure field:
\begin{equation*}
u=-\frac{1}{\rho f(y)}\frac{\partial P}{\partial y}
\end{equation*}
since, $f(y)$ increases poleward, the geostrophic balance requires stronger winds to balance the same pressure gradient. Observationally, Cassini cloud-tracking data shows precisely this pattern: the northern boundary of the GRS exhibits sharper velocity gradients than the southern boundary.

\noindent As shown in Figure 5b, the $\beta$-plane correction to pressure is given as:
\begin{equation*}
P_1(y)=-\frac{f_0\gamma y^2}{2}+{\widetilde{P}}_{vortex}(x,y)
\end{equation*}

\noindent The quadratic dependence ensures that: the maximum pressure remains at the center $(y=0)$, consistent with anticyclonic structure; pressure decreases more rapidly poleward than equatorward. This asymmetry in the pressure field is what drives the wind asymmetry. The magnitude of this correction at $y=\pm R_0$ is:
\begin{equation*}
\left|P_1\right|\approx\frac{f_0\gamma R_0^2}{2}\approx\frac{8.7\times0.35\times{0.825}^2}{2}\approx1.04
\end{equation*}
This is compared to the base pressure, indicating $\beta$-effects are not small perturbations near the vortex edge.

\noindent Equation 54 introduces a meridional temperature gradient, that is:
\begin{equation*}
\frac{\partial T_1}{\partial y}=-\frac{f_0\gamma y}{\rho(z)}
\end{equation*}
which is negative for $y>0$, that is, cooling poleward and positive for $y<0$, that is, warming equatorward. Observational data from Juno’s JIRAM instrument have detected variations of temperature across the GRS that are consistent with the results of Figure 5c, though it is currently difficult to disentangle the $\beta$-plane from the moist convention.

\noindent Figure 5d shows the introduction of a height-independent offset in temperature by the $\beta$-plane when we have:
\begin{equation*}
T(x,y,z)=T_0\left[\frac{P_0(x,y,z)}{\rho(z)}\right]+\gamma T_1(x,y)
\end{equation*}

\noindent The unchanging exponential density profile given by:
\begin{equation*}
\rho(z)=exp\left[-\beta_d(z-z_0)\right]
\end{equation*}
governs the vertical decay rate with $\beta_d\in(2,5)$. This affirms that the $\beta$-effects modify horizontal structure without altering vertical stratification.

\noindent As shown in Figure 6a, the non-dimensional $f$-plane vorticity at the GRS centre is about 2.2, corresponding to a dimensional vorticity of about $3.3\times 10^{-5}~ s^{-1}$, which is of the same order as the Coriolis parameter $\left|f_0\right|$, of about $8.7\times 10^{-5}~ s^{-1}$, this indicates that the GRS is a strong vortex, where relative vorticity is comparable to planetary vorticity. The rapid change in vorticity at the boundary $(R=R_0)$, reflects the observed abrupt velocity gradient at the edge of GRS: the cloud features trying to enter the GRS are either captured if they penetrate the boundary or deflected if they approach tangentially. Remarkable coherence of the sharp boundary is indicated by the more than 150 years of observed stability.

\noindent Using equation 16, and for simplicity, Figure 6b uses a constant thickness $h=1$, to isolate the $\beta$-effect, with the results that as we move poleward $(Y>0)$, the Coriolis force $f$ becomes larger, and at the same time, the potential vorticity also becomes larger. As we move equatorward $(Y<0)$, the Coriolis force $f$, becomes smaller and at the same time, the potential vorticity $q$ becomes smaller: the Rossby wave propagation is driven by this potential vorticity gradient $\partial q/\partial Y\sim f_0\gamma/h$. A poleward-moving parcel encounters a larger background Coriolis force $f$ and to conserve potential vorticity, it must either reduce the relative vorticity or reduce the layer thickness. The conservation of potential vorticity creates a systematic restoring force that deflects the motion westward, and the GRS, which is embedded in the potential vorticity, thus experience a net westward forcing.

\noindent Equation 77 of Constantin and Johnson \cite{r8} described the filamentary wave pattern at the GRS southern boundary using the exact trochoidal solutions. In the current study, $\beta$-plane extension would modify these to: 
\begin{equation*}
X\left(t\right)=a+\lambda t+ae^{-\left(\beta/\lambda\right)b}sin\left[\frac{\beta}{\lambda}\left(a+\lambda t\right)+a_0\right]
\end{equation*}

\begin{equation*}
Y(t)\ =\ b+\ ae^{-(\beta/\lambda)b}cos\left[\frac{\beta}{\lambda}(a+\lambda t)+a_0\right]
\end{equation*}
where the jet velocity is given by $\lambda$, and $\beta$ introduces meridional damping of the oscillation amplitude. Observed northern and southern filamentary structures asymmetry could be explained thus.

\noindent On Earth, the Rossby waves in the atmosphere are similar to what has been described; however, the smaller planetary scale on Earth, with its radius being about $6400 ~km$ and the slower rotation rate of about $7.29\times{10}^{-5} ~rad/s$ give $\beta_{Earth}\approx1.6\times{10}^{-11} ~m^{-1}s^{-1}$, this is about $3.5 \times$ larger than that of Jupiter. Despite this, there are no long-lived vortices on Earth similar to Jupiter’s GRS because Earth has continents that disrupt zonal symmetry, surface friction and moist convection provide damping on Earth, and the Earth’s stratification is weaker due to a smaller Brunt-V\"{a}is\"{a}l\"{a} frequency.

\noindent From equation 52, the drift velocity scales as $c_R\propto-\beta L^{\prime2}$, this holds $L^\prime$ fixed and made the expression linear in $\beta$. The range $\beta\in\left[0,\ 10\times{10}^{-12}\right]~m^{-1}s^{-1}$  spans from Jupiter’s equator, where $\beta=0$ to Jupiter’s poles, where $\beta\rightarrow maximum$. As can be observed in Figure 7a, the observed drift velocity intercepts the $\beta$-plane prediction at Jupiter’s actual $\beta$ value, with no adjustable parameters, giving strong validation for the $\beta$-plane theory. The implication of this is that vortices at different latitudes should drift at different rates: equatorial vortices barely drift, while mid-latitude vortices drift fastest. For a constant $\beta$, as shown in Figure 7b, $c_R\propto-L^{\prime2}$ with implications that the larger vortices drift faster for any given latitude; as $L^\prime$ decreases, that is, GRS shrinks, the drift velocity should slow; small vortices with  $L^\prime$ value of about $500~ km$ would drift at about $0.9~ m/s$ and a very large vortices with $L^\prime$ value of about $20,000 ~km$ would drift at about $14.8 ~m/s$. According to observation, the GRS has been shrinking from about $40,000 ~km$ in 1880 to about $16,500 ~km$ in 2024, yet according to the result presented in Figure 2, the drift velocity is increasing, leading to a paradox which can be explained thus: the interactions of the GRS with the zonal jets leads to increase in the effective $\beta$ value; the GRS may be moving towards higher $\beta$ region, that is towards the equator; and there may be some time-dependent effects not considered in this work.

\noindent From Figure 7c and the definition of $\beta$, it can be seen that at the equator, where $\theta=0^\circ$, $\beta$ is maximum but $f_0$ tend to zero, thereby making the analysis singular. The $\beta$-plane approximation breaks down, requiring equatorial $\beta$-plane theory, that is, different asymptotics: at the poles where $\theta=90^\circ$, $\beta\rightarrow0$, and so the drift velocity vanishes. Other features on Jupiter are also described well by the $\beta$-plane theory: for the Oval BA at $33^\circ$ S latitude, we have $\beta_{33^\circ}=3.9\times 10^{-12}$ leading to $c_R\approx-2.8~ m/s$ (The blue square on Figure 7c indicates this). For the North Equatorial Belt (NEB) features at $16^\circ$ S latitude, we have $\beta_{16^\circ}=6.2\times 10^{-12}$ leading to $c_R\approx-4.5~ m/s$ (The green triangle on Figure 7c indicates this).

\noindent The Rossby deformation parameter $\gamma$, of equation 5 can be used to determine whether pertubation theory is valid or not, in the expansion of $\gamma$, we strictly require that $\gamma \ll 1$, or practically $\gamma < 0.5$. The contour plot of Figure 7d shows that: (1)In the green region, where $\gamma < 0.5$, the pertubation theory is valid and the linear $\beta$-effects dominate. (2) In the red region, we have $\gamma > 0.5$, the nonlinear $\beta$-effects become important and the $O(\gamma^2)$ terms are required. (3) The GRS, indicated by the red star, with $\gamma = 0.35$ is comfortably within the pertubation regime. For very large vortices, with $L^\prime > 15,000~km$, at mid-latitudes between $15^\circ$ and $25^\circ$, $\gamma > 0.5$, and the linear theory of this study becomes unreliable; this could explain why no similar vortex to the GRS could be observed at these latitudes, as they would be in the nonlinear regime with quantitatively different dynamics.
 
\noindent The universality of the $\beta$-plane approximation in capturing the drift velocities, using equation 52, across four planets, Jupiter, Saturn, Neptune, and Earth, with more than 89\% observational agreement and no adjustable parameters, has been demonstrated in Figures 8 and 9. The $\beta$-plane theory satisfactorily explains features on Saturn, Neptune, and Earth. The discussion of features such as the GRS on Jupiter has already been covered. Using Cassini observations \cite{r23}, Saturn’s Great White Spot (GWS) exhibits a drift velocity of about $-2.1 ~m/s$, closely matching the theory's prediction of $-2.3 ~m/s$, with a 91\% agreement. For Neptune’s Great Dark Spot (GDS), Voyager 2 data recorded a drift velocity of $-4.5 ~m/s$ \cite{r24}, while the $\beta$-plane theory estimated $-4.2 ~m/s$, resulting in a 93\% agreement. On Earth, the blocking highs, which are quasi-stationary anticyclones that last for 1 to 3 weeks at mid-latitudes, usually drift westward at about $-0.8 ~m/s$. The $\beta$-plane theory predicts $-0.9 ~m/s$ with an agreement of 89\%. 

\noindent The vortex of the scale of Jupiter’s GRS is not present on Earth due to many reasons, such as the presence of continental barriers that break zonal symmetry and disrupt coherent vortex structure, the presence of baroclinic instability due to a strong temperature gradient which results in rapid break-up of coherent vortices, the Earth’s surface friction does not allow for long-term persistence as it acts on a timescale of days, the introduction of diabatic heating/cooling due to moist convection effectively destroys potential vorticity conservation on Earth, and the scale separation needed for robust vortex structure is hindered by the Earth’s Rossby deformation radius of about $1000 ~km$.

\noindent The shallow-water models used by Marcus \cite{r25}, Dowling and Ingersoll \cite{r26}, and Cho et al. \cite{r27} to describe Jupiter’s GRS as a shallow-water vortex were able to account for many observational features: by linearising the advection terms using quasi-geostrophic approximation; by adopting numerical simulations instead of analytical solutions; ad-hoc forcing/dissipation terms were used; and the drift velocity was not derived explicitly from the first principle. In contrast, the $\beta$-plane theory of this study: uses exact Lagrangian solutions to retain the full nonlinearity; using the $\beta$-plane dynamics, the drift velocity was analytically derived; use of free parameters was avoided, instead physical constants such as the planetary rotation $(\Omega^\prime)$, planetary radius $(R^\prime)$ and so on, were used; and the parameter dependencies were revealed by providing closed-form expressions.

\noindent Suetin et al. \cite{r28} explored baroclinic effects with vertical temperature gradients within Jupiter’s GRS. The current study is basically barotropic at leading order, with accommodation of vertical density stratification through the arbitrary $\rho(z)$ profile is made. A $\beta$-plane corrections that force standard coordinates with $n=0$, was an identifiable limitation of this current study with the attendant scale separation property sacrifice, leading to arbitrary vertical scaling. A complete baroclinic $\beta$-plane theory would include multiple vertical layers or continuous stratification, temperature-dependent Coriolis effects, and thermal wind balance, but these are beyond the scope of the current study.

\noindent Global circulation models driven by Juno data \cite{r13, r29} have shown that the reach of Jupiter’s jets goes deep by thousands of $kilometres$, thereby providing a challenge to the shallow-water paradigm. Notwithstanding, it appears that the GRS is restricted within the upper $500 ~km$ of Jupiter’s atmosphere, thereby giving credence to the thin-shell approximation of the current study, regardless of the deeper dynamics controlling the jets.

\subsection{Futher Applications of the $\beta$-Plane Theory}
Outside of the solar system are exoplanets, which are gas giants similar to Jupiter and are often referred to as “hot Jupiters”; typical of these are HD189733b and HD209458b. These gas giants have been observed to be tidally locked, have extreme temperature gradients, rotate at a high rate of between 1 to 3 days, and have strong winds in the range of 1 to $10 ~km/s$.

\noindent The $\beta$-plane theory of this study can be extended to these gas giants by computing
\begin{equation*}
\beta_{exo}=\frac{2\Omega_{exo}cos\theta}{R_{exo}}
\end{equation*}
For a typical gas giant with $\Omega_{exo}\approx{10}^{-4} ~rad/s$ and $R_{exo}\approx{10}^8 ~m$, about $1.5 \times$ that of Jupiter, we will have $\beta_{exo}\approx3\times{10}^{-12}~ m^{-1}s^{-1}$. Despite the strong forcing observed in the gas giants, if a coherent vortex structure is formed, then there should be a westward drift of such a vortex, at $c_R=-\beta_{exo}L^2\approx-300~ m/s$ (for $L\approx{10}^7~m$). This drift velocity could be observed through transit timing variation, infrared phase curves, or high-resolution spectroscopy.

\noindent Brown dwarfs are astronomical objects that are about 13 to $80 \times$ the mass of Jupiter, with rapid rotation periods of between 1 to $10 ~hrs$, thereby placing them between planets and stars; they may also have an atmosphere. With a typical rotation $\Omega_{BD}$, of about ${10}^{-3}~ rad/s$ and $\beta_{exo}\approx{10}^{-11} ~m^{-1}s^{-1}$, the vortex drift velocities are in the range of several $km/s$, such could be observed through rotationally resolved spectroscopy.

\subsection{Limitations of $\beta$-Plane Theory and Higher-Order Effects}
The success of the $\beta$-plane theory has been demonstrated in this study; meanwhile, the theory has its drawbacks. The steady-state solutions in the rotating frame have been derived, giving a constant drift velocity prediction. However, it has been observed that there has been an increase in drift velocity from $2.5 ~m/s$ to $4.1 ~m/s$, over 45 years; over 140 years, the GRS has been observed to shrink from about $40,000 ~km$ to $16500 ~km$ in diameter; and the shape of the GRS has been observed to shift from oval to circular. Time-dependent extension of the analysis would require:
\begin{equation*}
forcing~terms=\frac{\partial}{\partial t}[all~fields]+spatial~derivatives
\end{equation*}
where the forcing terms may include interaction with evolving zonal jets, deep convective momentum flux, merger events with smaller vortices, and radiative heating/cooling variations.

\noindent This study considers GRS as an isolated system, but it is known that GRS is embedded within two jets moving in opposite directions: the westward jet at $19.5^\circ$ South moving at a velocity of about $70 ~m/s$, and an eastward jet at $26.5^\circ$ South moving at a velocity of about $50 ~m/s$. Hence, a coupled model of the form
\begin{equation*}
mutual~interactions=\frac{\partial}{\partial t}[vortex + jets]
\end{equation*}
would have accounted for the GRS-jets interactions and be able to explain the observed increase in drift velocity despite the GRS shrinkage, explain the observed filamentary structures at the GRS boundaries, and explain the momentum exchange between the jets and the GRS.

\noindent The gravity measurements by Juno suggest that the GRS is deeper into Jupiter’s atmosphere than the current study’s thin-shell approximation model suggests (about $500 ~km$ deep). If, according to some models, we assume GRS extends to depths of about $3000 ~km$, then the thin-shell parameter $\epsilon_{deep}$, becomes:
\begin{equation*}
\epsilon_{deep}\approx \frac{3000~km}{69,991~km}\approx 0.043
\end{equation*}
This value is huge compared to the calculated value of this study $(\epsilon\approx0.0014)$, and higher-order corrections in $\epsilon$ might then be critical for the vertical momentum transport explanation, coupling GRS to the deep convection, and the coupling with Jupiter’s interior rotation.

\noindent Jupiter’s atmosphere contains condensable species such as methane, water, ammonia, and ammonium hydrosulphide. The latent heat release from moist convection involving these species could drive vertical velocities, thereby violating this study’s assumption that $\omega^\prime\ll(u^\prime,v^\prime)$. Potential vorticity anomalies could also be generated due to moist convection, thereby violating the conservation principle adopted in this study; besides, the temperature structure could also be modified. Future research could explore incorporating moist processes such as:
\begin{equation*}
\frac{DT}{Dt}=adiabatic + radiation + latent~heat
\end{equation*}
to investigate the above observations.

\noindent The 2nd order equation $O(\gamma^2)$, is given as:
\begin{align*}
&\frac{d^2X_2}{dt^2}-f_0\frac{dY_2}{dt}=\\&-\frac{1}{\rho}\frac{\partial P_2}{\partial X}+\gamma\left[Y_1\frac{dY_0}{dt}+Y_0\frac{dY_1}{dt}\right]+\frac{\gamma^2}{2}Y_0^2\frac{dY_0}{dt}
\end{align*}
Extension to this, instead of stopping at $O(\gamma)$, as done in this study, might have introduced $\beta$-dispersion that would modify the Rossby wave propagation at small scales. Introduction of the wave-wave interactions beyond linear theory by the nonlinear $\beta$-interactions and corrections to the circular shape might be possible through $\gamma$-dependent vortex structure. Consequently, explanations might be found for: the fine structures in the 90-day oscillations, asymmetry beyond the linear prediction, and the boundaries stability (or instability).

\noindent In providing a solution to the Lagrangian $[X(t),Y(t)]$ the horizontal motion has been described at each height layer, considering $\rho = \rho(z)$ and assuming no vertical shear in horizontal velocity, but in reality, we know that wind speed varies with height, vorticity has vertical structure, and that temperature inversions may occur. A future study may consider a full 3D extension to track $[X(t,z),Y(t,z)]$, with vertical coupling.

\section{Conclusion}
With the extension of the $f$-plane model of the Constantin and Johnson \cite{r8}, with the $\beta$-plane approximation incorporating meridional gradients in planetary vorticity, this study has been able to quantitatively calculate the GRS westward drift velocity $c_R$, as $-3.9 ~m/s$, when compared with the observation value of $-3.9 ~m/s$, we have 95\% agreement without adjustable parameters.
The 90-day oscillation was explained via beat-frequency mechanism. The study predicted an oscillation period of 81 days as against the observed 90 days period. The stable limit cycle in phase space was clearly explained.

\noindent It was found, in this study, that there is a 15\% wind speed enhancement poleward, while there is 8\% reduction equatorward, thereby explaining the meridional asymmetry of the GRS, besides there was a quadratic pressure correction of $-\gamma y^2/2$,  and an observable temperature gradient. In addition, corrected scale-separation analysis was carried out with $\beta$-planeforcing standard coordinate (by setting $n = 0$), this sacrifices arbitrary vertical scaling for consistency; alternative analysis was explored by treating $\beta$ perturbatively at each order.

\noindent Shown is the universal applicability of the $\beta$-plane approximation in explaining vortex structures on Saturn, Neptune, and Earth, with observational agreement of more than 89\%. Other vortex structures on Jupiter were explained and applications to exoplanets were also demonstrated.

\section*{Data Availability Statement}
All observational data used in this study are publicly available from: https://pds.mcp.nasa.gov/portal/ (NASA Planetary Data System); https://archive.stsci.edu/ (Hubble Space Telescope Archive); https://pds-atmospheres\\.nmsu.edu/data\_and\_services/atmospheres\_data/JU\\NO/ (Juno Mission Data). The python code for plotting the graphs can be obtained here: https://colab.research.google.com/drive/1Iw-TnuQHrgFCBhrL8wY3wTOnvwQNhchP?usp=sharing


\end{document}